\documentclass[twocolumn]{aastex7}

\usepackage{graphicx,caption, subcaption}
\usepackage{amsmath}
\usepackage{pifont}

\begin{document}

\title{The Plane Quasar Survey: An Ionized Extension of the Magellanic Stream on the Northern Side of the Galactic Plane}

\author[0000-0002-9040-672X, sname='Choi']{Bo-Eun Choi} 
\affiliation{Department of Astronomy, University of Washington, 
Seattle, WA 98195, USA}
\email[show]{bechoi@uw.edu}

\author[0000-0002-0355-0134, sname='Werk']{Jessica K. Werk}
\affiliation{Department of Astronomy, University of Washington, 
Seattle, WA 98195, USA}
\email[show]{jwerk@uw.edu}

\author[0000-0003-0789-9939]{Kirill Tchernyshyov}
\affiliation{Department of Astronomy, University of Washington, 
Seattle, WA 98195, USA}
\email[]{ktcherny@gmail.com}

\author[0000-0002-1129-1873]{Mary E. Putman}
\affiliation{Department of Astronomy, Columbia University, New York, NY 10027, USA}
\email[]{mputman@astro.columbia.edu}

\author[0000-0003-4158-5116]{Yong Zheng}
\affiliation{Department of Physics, Applied Physics and Astronomy, Rensselaer Polytechnic Institute, Troy, NY 12180, USA}
\email[]{zhengy14@rpi.edu}

\author[0000-0003-4797-7030]{J. E. G. Peek}
\affiliation{Space Telescope Science Institute, 3700 San Martin Drive, Baltimore, MD 21218, USA}
\email[]{jegpeek@stsci.edu}

\author[0000-0002-7483-8688]{Hannah Bish}
\affiliation{Space Telescope Science Institute, 3700 San Martin Drive, Baltimore, MD 21218, USA}
\email[]{hbish@stsci.edu}

\author[0000-0003-2666-4430]{David Schiminovich}
\affiliation{Department of Astronomy, Columbia University, New York, NY 10027, USA}
\email[]{ds@astro.columbia.edu}




\begin{abstract}
The Magellanic Stream (MS) is a vast gaseous structure in the Milky Way halo, containing most of its mass in ionized form and tracing the interaction between the Large and Small Magellanic Clouds and the Galaxy. 
Using HST/COS G160M spectra from the Plane Quasar Survey, we detect \ion{C}{4} absorbers likely associated with the MS, extending to the northern side of the Galactic plane, approximately 60$^\circ$ beyond its previously known ionized extent. These absorbers exhibit position and kinematic alignment and show consistent ionization trends with previously studied MS sight lines. The non-detection of low ions such as \ion{Al}{2} and \ion{Si}{2}, and the detection of \ion{C}{4} (and \ion{Si}{4} in some sightlines), indicates a highly ionized gas phase. 
The observed \ion{Si}{4}/\ion{C}{4} column density ratios suggest a gas temperature of $T \sim 10^{5.3}$~K and favor collisional ionization over photoionization. We estimate the newly detected extension increases the previous ionized gas mass of the MS, and its coherent kinematics suggest that it was stripped within the past few hundred Myr and has not yet mixed with the Milky Way halo. The existence of highly-ionized MS gas at a location above the Galactic Plane may constrain the orbital direction of the Magellanic Clouds. 
\end{abstract}




\section{Introduction} \label{sec:intro}

The Magellanic Stream (MS; hereafter the Stream) is a trailing gaseous structure that traces the interaction of the Large and Small Magellanic Clouds (LMC and SMC; hereafter, the ``Clouds") with the Milky Way (MW). 
Its full extent in \ion{H}{1} 21~cm emission spans approximately 150~degrees across the southern Galactic sky, with a width of 10~degrees \citep{mathewson74, putman03, nidever10}. 
The Stream also represents the most massive neutral high-velocity cloud (HVC) complex in the MW halo \citep{putman12}, with a \ion{H}{1} mass of $M_{\rm HI} \approx 2.7 \times 10^8\ M_{\odot}$ \citep{bruns05}. 
The Stream represents the single most massive and extended gaseous structure in the MW halo, serves as a testbed for studying the impact of satellite galaxy interactions, and is a crucial probe of coherent gaseous circumgalactic structures throughout the universe.

Observational studies of the neutral Stream, using deep \ion{H}{1} surveys and UV absorption spectroscopy, have revealed that the Stream consists of two filaments with distinct morphologies, kinematics \citep{putman03, nidever08}, and metallicities \citep{fox13, ritcher13}. These findings suggest that the Stream contains material stripped from both the Clouds; however, its detailed formation history and nature are not yet fully understood. 
As the position-velocity structure of the Stream serves as a key observational constraint on the Clouds' orbital history and the nature of its formation, many theoretical models have focused on reproducing it. 
The Stream was traditionally explained by tidal stripping of material from the Clouds during multiple passages around the MW \citep[e.g.,][]{lin77}, with later models including the LMC-SMC interactions (i.e., second passage; \citealt{yoshizawa03, diaz12}). 
Precise \textit{Hubble Space Telescope} proper motion measurements of the Clouds posed a different view of their orbital history, suggesting that they may have completed only one orbit or may even be on their first passage around the MW \citep{kallivayalil06a, kallivayalil13}. This led to the development of first-passage models, in which the Stream forms primarily through tidal interactions between the Clouds \citep{besla10, besla12}.

Beyond its neutral gas component, the Stream also hosts a significant reservoir of ionized gas, which introduces additional complexity in understanding its formation and evolution. \cite{fox14} estimated the total ionized mass of the Stream to be $\sim 10^9\ M_{\odot}$, exceeding the neutral mass and extending more widely across the sky. 
This dominant mass in ionized gas is underestimated in tidal models \citep{diaz12, besla12}. 
As an alternative to tidal models, ram-pressure models have been suggested in which the Stream is formed via ram-pressure stripping of gas from the Clouds while they pass through the extended, gaseous MW halo \citep[e.g.,][]{mastropietro05, hammer15}. 
\cite{wang19} applied a ram-pressure plus direct collision model and reproduced the observed properties including the ionized and \ion{H}{1} gas mass. 
\cite{lucchini20} proposed that the Magellanic corona, a warm, extended gas envelope surrounding the Clouds, plays a crucial role in the formation of the Stream, contributing a substantial fraction of its ionized gas. 
Recent UV detections provide some observational evidence of a $10^{5.5}$~K corona, however it is difficult to distinquish a corona from the ionized component of the Stream \citep{krishnarao22, kim24}.

The ionization and kinematic structure of the ionized gas in the Stream are crucial for understanding the physical processes that shapes its evolution. 
UV absorption observations show that the ionization fraction increases toward the trailing end of the Stream \citep{kim24}. 
The underlying ionization mechanisms remain uncertain \citep{putman03b}, with potential sources including shock ionization as the Stream travels through the MW halo \citep{weiner96, bland-hawthorn07}, a recent Seyfert flare from the Galactic Center \citep{bland-hawthorn19}, and photoionization by stellar radiation from the MW and Clouds \citep{bland99, barger17}. 
In addition to the ionization structure, the velocity structure of the ionized gas provides a useful tracer of the Stream’s interaction with the CGM of the MW and LMC, as the more diffuse ionized component is more sensitive to hydrodynamic drag \citep{zhu24, lucchini24b} and likely traces conductive or turbulent boundary layers surrounding the Stream \citep{bustard22,fox05}.

In this Letter, we report the serendipitous discovery of an ionized extension of the Stream reaching above the Galactic plane, about 60~degrees longer than the previously detected ionized gas. This ionized component is traced by \ion{C}{4} absorption in QSO sight lines near the Galactic plane. The QSOs were selected as part of the Plane Quasar Survey (PQS; \citealt{werk24}), which was originally designed to probe gas structures near the MW’s disk unrelated to the Stream. 
From the observations with the \textit{Hubble Space Telescope} (HST)/Cosmic Origin Spectrograph (COS), we detect \ion{C}{4} absorbers with kinematics consistent with those of the Stream up to the Galactic latitude of $b \sim 27^\circ$. 
We describe the HST/COS observation and data analysis in Section~\ref{sec:data}. 
In Section~\ref{sec:results}, we discuss the association of newly detected \ion{C}{4} absorbers with the Stream, presenting their velocity structure, ionization states, and estimated mass. We discuss their origin and the physical implications of our discovery in Section~\ref{sec:discussion}. 

Throughout this manuscript, all velocities are reported in the local standard of rest (LSR) frame, denoted as $v_{\rm LSR}$. 
We adopt the barycentric solar motion relative to the LSR as $(U, V, W) = (11.1, 12.24, 7.25)\ \rm km\ s^{-1}$ \citep{schonrich10}. The LSR frame is commonly used in studies of HVCs, enabling direct comparison with previous observations and models of the Stream and other halo gas structures.

\section{Observation \& Data Analysis} \label{sec:data}

\subsection{HST/COS G160M Observation} \label{subsec:observation}

\begin{figure*}[ht!]
  \centering
  \begin{subfigure}[t]{\textwidth}
    \centering
    \includegraphics[width=\linewidth]{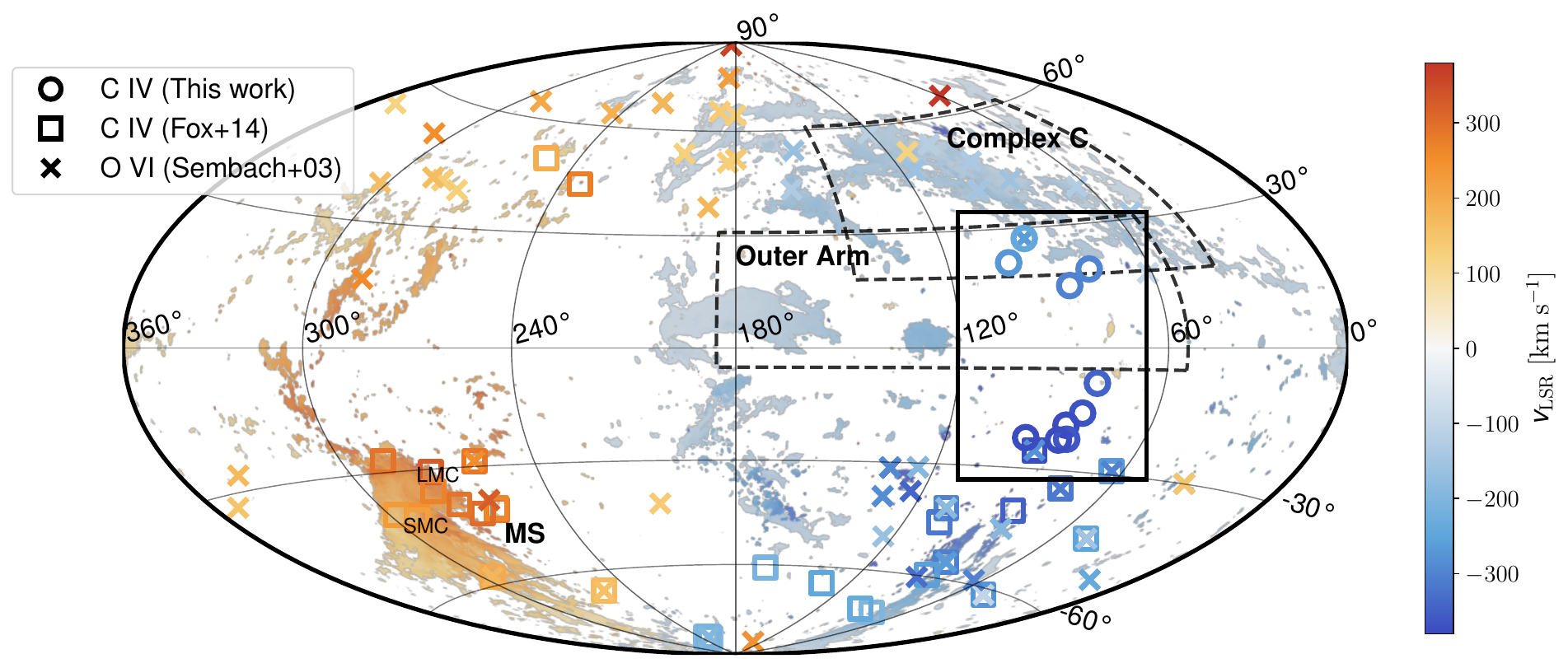}
    \caption{The location of high-ion absorbers near the Stream in Galactic coordinates.}
    \label{fig1:gal_dist}
  \end{subfigure}
  \quad
  \begin{subfigure}[b]{\textwidth}
    \centering
    \includegraphics[width=\linewidth]{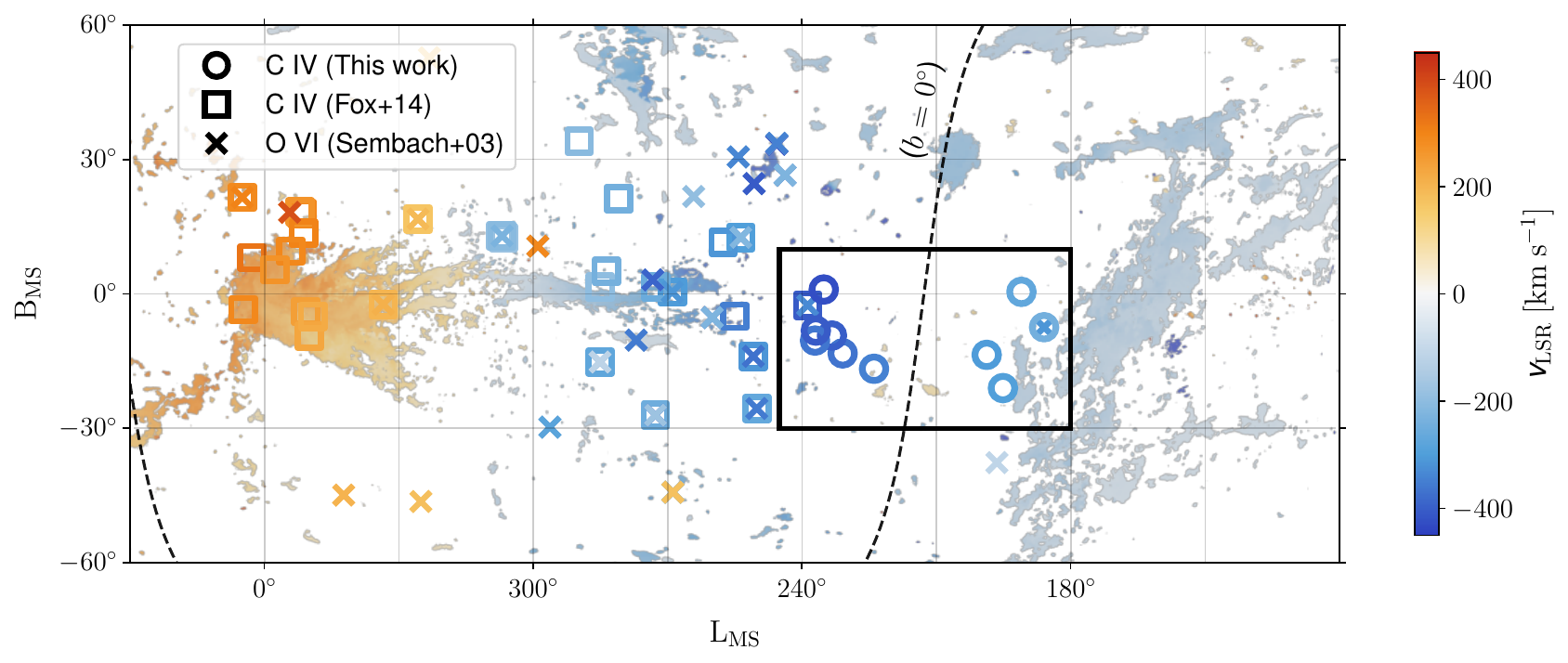}
    \caption{The location of high-ion absorbers near the Stream in MS coordinates.}
    \label{fig1:MS_dist}
  \end{subfigure}
  \caption{The location of high-ion absorbers near the Stream in (a, top panel) Galactic coordinates and (b, bottom panel) MS coordinates. In this work, we detect \ion{C}{4} absorbers (circles), extending to the Northern sky (Galactic latitude of $b \sim 30^{\circ}$). We also include \ion{C}{4} (squares; \citealt{fox14}) and \ion{O}{6} (x markers; \citealt{sembach03}) absorbers from previous studies. The LSR velocity of the absorbers is color-coded. The all-sky \ion{H}{1} map from the HI4PI survey \citep{westmeier18} is shown with the same velocity color scale.}
  \label{fig:targets_dist}
\end{figure*}

\begin{deluxetable*}{lcccccccc}
    \tablecaption{Data Summary \label{tab:data}}
    \tablehead{\colhead{QSO Name} & \colhead{$l$} & \colhead{$b$} & \colhead{$L_{\rm MS}$} & \colhead{$B_{\rm MS}$} & \colhead{Instrument} & \colhead{$\log{N_{\rm CIV}}$} & 
    \colhead{$b_{\rm CIV}$} & \colhead{$v_{\rm LSR, CIV}$} \\
    \colhead{} & \colhead{(deg)} & \colhead{(deg)} & \colhead{(deg)} & \colhead{(deg)} & \colhead{} & \colhead{($\rm cm^{-2}$)} & \colhead{($\rm km\ s^{-1}$)} & \colhead{($\rm km\ s^{-1}$)}
    }
    \startdata
    J2206+2757 & 84.0 & -22.2 & -123.0 & -10.4 & COS/G160M & $13.78 \pm 0.04$ & $45.0 \pm 5.7$ & $-345.8 \pm 3.9$ \\
    J2215+2902 & 86.4 & -22.5  & -123.1 & -8.2 & COS/G160M & $13.50 \pm 0.09$ & $30.0 \pm 8.9$ & $-388.8 \pm 5.8$ \\
    J2251+3419 & 96.5  & -22.3  & -124.9 & 1.0 & COS/G160M & $13.73 \pm 0.05$ & $41.1 \pm 6.9$ & $-412.7 \pm 4.8$ \\
    J2203+3145 & 86.0    & -18.8  & -126.7 & -9.3 & COS/G160M & $13.89 \pm 0.03$ & $48.0 \pm 4.2$ & $-383.2 \pm 2.9$ \\
    J2141+3151 & 82.4  & -15.7  & -129.1 & -13.3 & COS/G160M & $14.16 \pm 0.04$ & $53.6 \pm 5.8$ & $-343.7 \pm 4.1$ \\
    J2109+3532 & 80.3  & -8.4   & -136.1 & -16.7 & COS/G160M & $13.90 \pm 0.07$ & $30.0 \pm 6.3$ & $-321.8 \pm 4.5$ \\
    J1938+5408 & 86.4  & 15.3   & -161.3 & -13.6 & COS/G160M & $13.37 \pm 0.03$ & $25.3 \pm 3.0$ & $-273.9 \pm 1.9$ \\
    J1858+4850 & 78.8  & 19.1   & -164.9 & -21.0 & COS/G160M & $13.35 \pm 0.09$ & $11.3 \pm 6.1$  & $-265.4 \pm 3.2$ \\
    J1939+7007 & 102.0   & 21.5   & -169.0 & 0.5 & COS/G160M & $13.02 \pm 0.14$ & $10.0 \pm 10.3$ & $-247.8 \pm 5.1$ \\
    H1821+643  & 94.0 & 27.4  & -174.1 & -7.4 & STIS/E140M & $13.31 \pm 0.04$ & $6.6 \pm 0.9$ & $-216.2 \pm 0.6$ 
    \enddata
    \tablecomments{Properties of the targeted QSO sight lines, including their names, Galactic coordinates ($l$, $b$), MS coordinates ($L_{\rm MS}$, $B_{\rm MS}$), and observed instruments. The best Voigt profile fitting parameters to the \ion{C}{4} MS absorbers, \ion{C}{4} column density ($N_{\rm CIV}$), $b$-parameter, and velocity centroid in the LSR, are also presented.}
\end{deluxetable*}

We originally proposed observations of the PQS targets to investigate the extent of a rotating, extended, ionized disk of the Milky Way, with selected QSO sight lines from the PQS at $l \approx 90^{\circ}$ and $\approx 270^{\circ}$ at low Galactic latitudes of $|b| < 30^{\circ}$ (Fig.~\ref{fig1:gal_dist}). 
We observed the QSOs and obtained their FUV spectra with the HST/COS G160M grating with a central wavelength of 1577~\AA\ (Program ID: 16679). 
Exposure times were optimized to achieve a signal-to-noise ratio (S/N) $\sim$ 11-15 per resolution element across 1400-1550~\AA. 
The COS G160M grating provides a wavelength coverage of 1360-1775~\AA, covering absorption lines from various metal ions, including low ions (e.g., \ion{Si}{2}$\lambda$1526, \ion{Al}{2}$\lambda$1670) and intermediate ions (i.e., \ion{Si}{4}$\lambda\lambda$1393, 1402, \ion{C}{4}$\lambda\lambda$1548, 1550). 
A characteristic spectral resolution of the G160M grating at lifetime position 6 (LP6) is $R \sim 10,000$. 

We applied \texttt{coadd\_x1d} routine developed by \cite{danforth10, danforth16} to the CALCOS-generated \texttt{x1d.fits} files. It combines all the exposures and treats the error arrays properly based on Poisson statistics. 
We combined the exposures with a weight on S/N and binned the data by three spectral pixels. The resulting science-grade spectra have an FWHM $\approx 18\ \rm km\ s^{-1}$.

A useful coordinate system to examine the structure associated with the Stream is the MS coordinate frame defined by \cite{nidever08}. We show the locations of our targeted sight lines in MS coordinates, $(L_{\rm MS}, B_{\rm MS})$, in Fig.~\ref{fig1:MS_dist}. 
In addition to our COS G160M observations, we searched the Mikulski Archive for Space Telescopes (MAST) for FUV QSO spectra within $-180^\circ < L_{\rm MS} < -150^\circ$ and $|B_{\rm MS}| < 30^\circ$, and obtained an HST/STIS E140M spectrum of H1821+643 (Program ID: 8165). 
Table~\ref{tab:data} contains a summary of our QSO samples. 
The specific observations analyzed in this article can be accessed via \dataset[doi: 10.17909/78fe-ne69]{https://doi.org/10.17909/78fe-ne69}.

\subsection{Data Analysis} \label{subsec:data analysis}
\begin{figure*}[ht!]
    \centering
    \includegraphics[height=0.95\textheight]{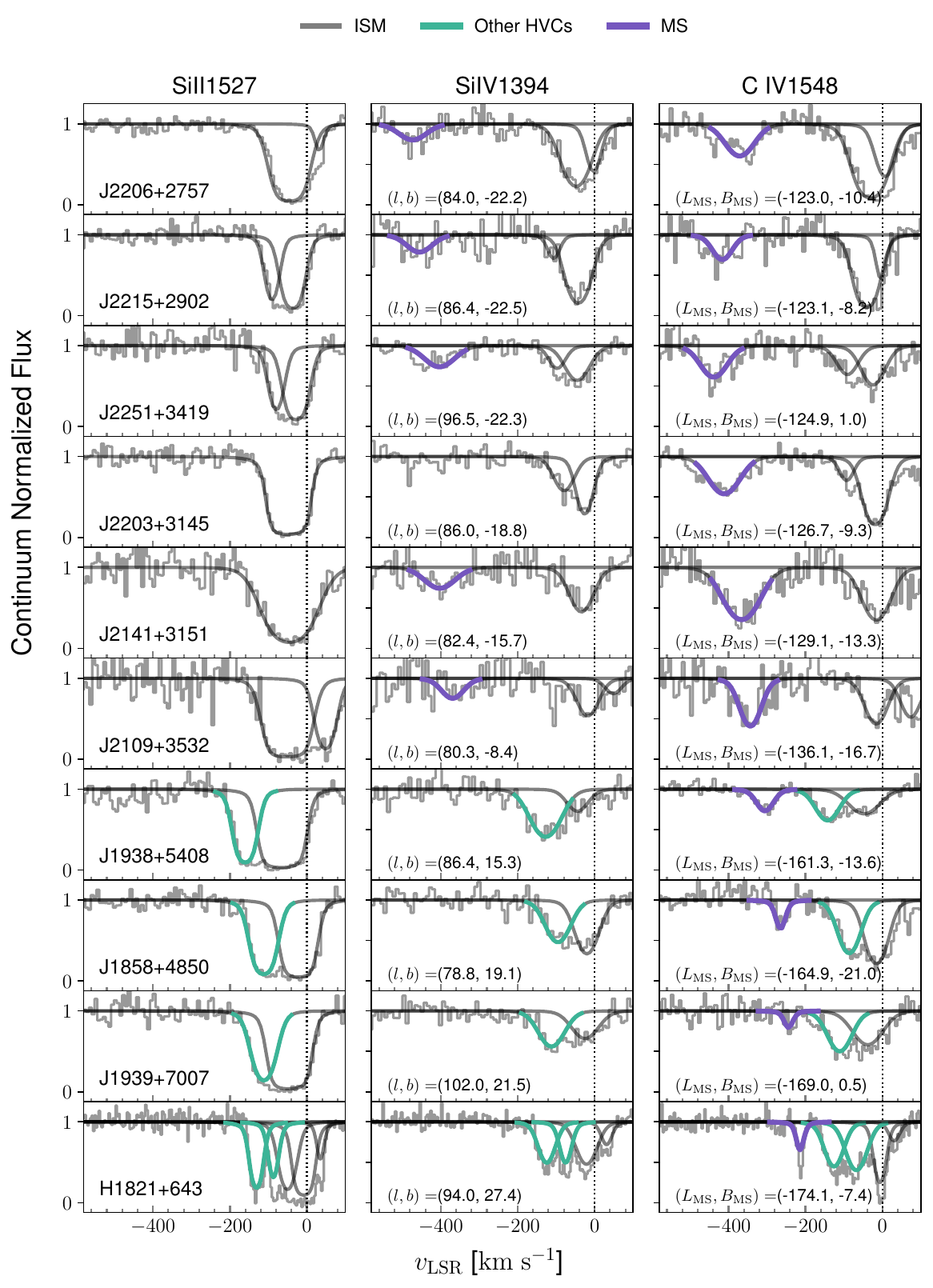}
    \caption{UV spectra and fitted Voigt profiles for \ion{Si}{2} $\lambda1527$ (left), \ion{Si}{4} $\lambda1394$ (center), and \ion{C}{4} $\lambda1548$ (right) absorptions in the QSO sight lines. The components associated with the MS are shown in purple. These components have distinct velocity offsets from other HVCs, such as Complex~C (green), and are not detected in low ions like \ion{Si}{2}. 
    \label{fig:vpf}}
\end{figure*}

We normalize the continuum of all spectra using \texttt{lt\_continuumfit}, a Python tool in the open-source package \texttt{linetools} \citep{prochaska17a}. 
We identify all the absorption components in LSR velocity space within $|v_{\rm LSR}| < 500\ \rm km\ s^{-1}$. 
The LSR velocity window is chosen to detect absorption components associated with the Stream, of which known LSR velocity of neutral gas spans $|v_{\rm LSR}| \lesssim 450\ \rm km\ s^{-1}$, as well as with the MW ISM and other HVCs. 
We use the \texttt{pyigm IGMGuesses} \citep{pyigm} to identify components by eye using offsets of velocity centroids. 
We then jointly fit Voigt profiles to every absorption component in the velocity window using the public package \texttt{veeper\footnote{The original version of \texttt{veeper}: \url{https://github.com/jnburchett/veeper}. 
In this work, we use a custom version (\url{https://github.com/mattcwilde/veeper}).}}. 
We present the best-fit Voigt profile parameters, ion column density ($N_{\rm ion}$), Doppler-$b$ parameter, and velocity centroid ($v_{\rm LSR}$), for all the sight lines and all the detected metal ions in Table~\ref{tab:all_VPF} in Appendix~\ref{appendix:vpf}.

\section{Results} \label{sec:results}

\begin{figure*}[ht!]
    \includegraphics[width=\linewidth]{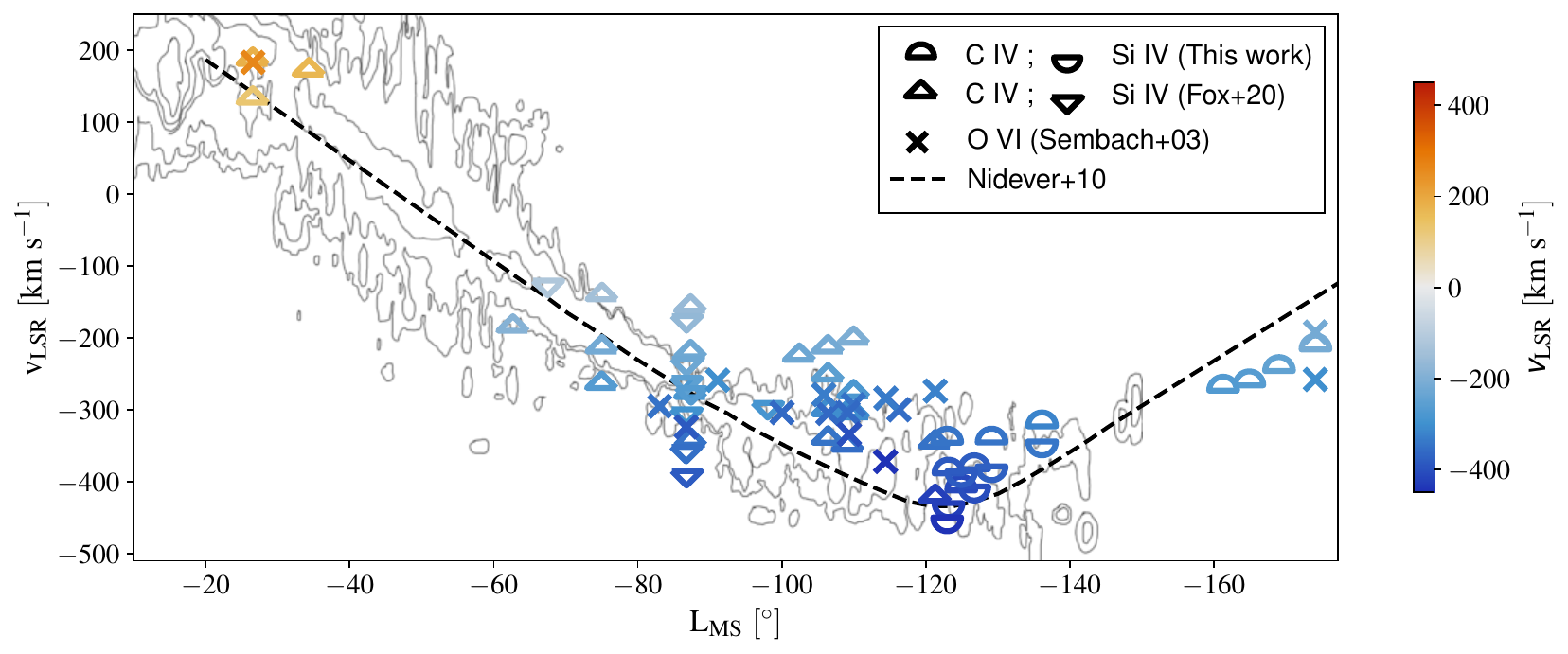}
    \caption{Velocity centroid of \ion{C}{4} (half-top circles) and \ion{Si}{4} (half-bottom circles) MS absorbers in the LSR by MS Longitude ($L_{\rm MS}$). The \ion{C}{4} and \ion{Si}{4} from \cite{fox20} are also shown in half-top and half-bottom diamonds, respectively, and \ion{O}{6} absorbers from \cite{sembach03} are shown as x marks. 
    The velocity gradient suggested by \cite{nidever10} is presented with dashed line, and our data follows this trend including velocity inflection at $L_{\rm MS} \sim -120^{\circ}$. 
    The background \ion{H}{1} contours are adopted from \cite{nidever10} with permission.
    \label{fig:vlsr_L_MS}}
\end{figure*}

\begin{figure*}[ht!]
    \includegraphics[width=\linewidth]{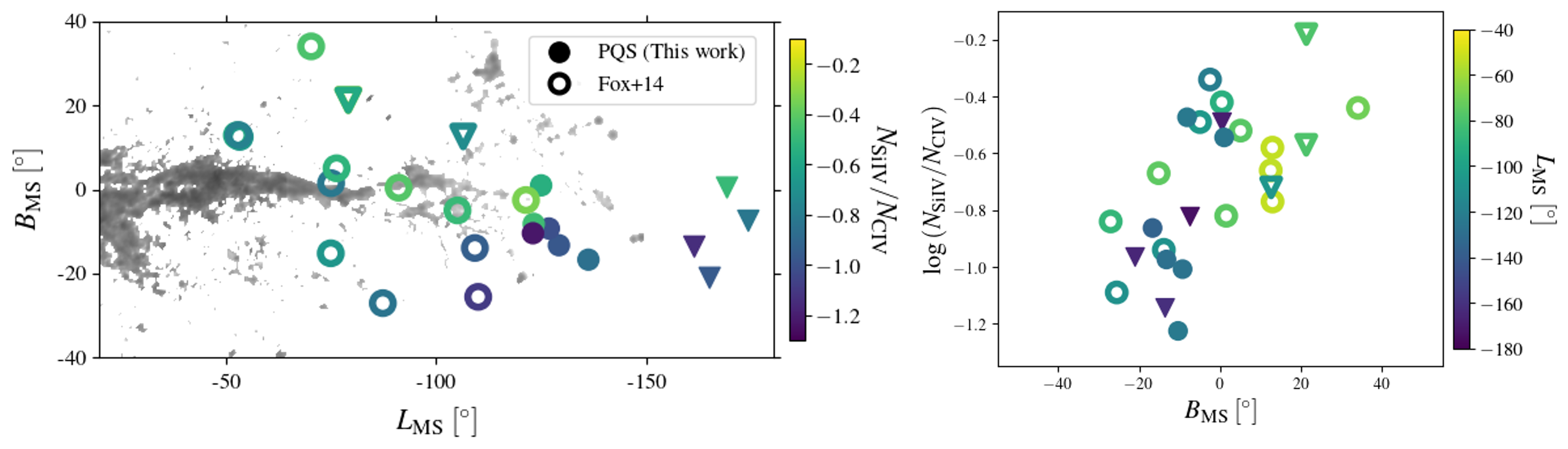}
    \caption{The ion column density ratios of \ion{Si}{4} and \ion{C}{4} along the Stream. (Left) QSO sight lines with detected \ion{C}{4} absorption, shown in MS coordinates and color-coded by \ion{Si}{4}/\ion{C}{4} column density ratio. Filled circles represent measurements from this work, and open circles are from \citet{fox14}. Triangles denote sight lines where only upper limits on \ion{Si}{4} are available. 
    (Right) \ion{Si}{4}/\ion{C}{4} column density ratios along with MS Latitude ($B_{\rm MS}$). 
    \label{fig:SiIV_CIV}}
\end{figure*}

In Fig.~\ref{fig:vpf}, we present the FUV spectra with best-fit Voigt profiles for \ion{Si}{2}$\lambda1527$ (left), \ion{Si}{4}$\lambda1394$ (center), and \ion{C}{4}$\lambda1548$ (right). 
The Voigt profiles for all absorption components within $|v_{\rm LSR}| < 500\ \rm km\ s^{-1}$ are shown in black, while components associated with the Stream are highlighted in purple. Hereafter, we refer to these components as ``MS absorbers". Other high-velocity absorption features associated with known structures are shown in green. 
We detect \ion{C}{4} MS absorbers in 10 sight lines. No corresponding \ion{Si}{2}$\lambda1527$ absorption is detected, nor is absorption from other low-ion species such as \ion{Al}{2}$\lambda1670$ (Table~\ref{tab:all_VPF}). 
In 5 sight lines with relatively higher \ion{C}{4} column densities, \ion{Si}{4} is also detected. 
The detection of only intermediate ions, particularly \ion{C}{4}, implies that the gas in this region is highly ionized.

\subsection{\ion{C}{4} Absorbers Associated with the Stream} \label{subsec: association}

We briefly summarize the main arguments supporting the association of the detected \ion{C}{4} absorbers with the Stream:
\begin{itemize}
    \item \textbf{Kinematic alignment with Stream predictions.} Velocity centroids align with the velocity gradient of the Stream derived from \ion{H}{1} gas at $L_{\rm MS} \gtrsim -150^\circ$ \citep{nidever10}, while our MS absorbers extend further to $L_{\rm MS} \sim -180^\circ$ (Fig.~\ref{fig:vlsr_L_MS}). 
    \item \textbf{Consistent ionization trend with known Stream sight lines.} Column density ratios of \ion{Si}{4}/\ion{C}{4} are consistent with those found in sight lines near the traditional end of the Stream, following a similar trend with MS latitude (Fig.~\ref{fig:SiIV_CIV}). 
    \item \textbf{Distinct from known Galactic \ion{H}{1} structures in the disk and halo.} The absorbers differ from Complex~C and the Outer Arm in ionization (only \ion{C}{4} detected), show offset in velocity (more negative $v_{\rm LSR}$), and do not follow the velocity trends of these known Galactic \ion{H}{1} structures. 
\end{itemize}

We examine the association between these MS absorbers and the Stream in detail. 
The LSR velocity of neutral gas in the Stream has been well characterized through \ion{H}{1} 21~cm observations \citep{putman03, bruns05, nidever10}, making it a useful reference for identifying absorbers associated with the Stream. 
The Stream exhibits a velocity gradient along the MS longitude ($L_{\rm MS}$), with $v_{\rm LSR}$ decreasing nearly linearly and reaching $-440\ \rm km\ s^{-1}$ at the far end of the neutral Stream ($L_{\rm MS} \sim -120^{\circ}$). This velocity gradient, which provides a critical constraint for formation models of the Stream, has been reproduced in recent simulations (e.g., \citealt{besla12, diaz12, lucchini20, lucchini21}). 
Meanwhile, \cite{nidever10} detected \ion{H}{1} emission extending to $L_{\rm MS} \sim -150^\circ$ in deep \ion{H}{1} 21~cm observations. 
Their analysis of this extended region revealed a velocity inflection point near $L_{\rm MS} \sim -120^{\circ}$, where $v_{\rm LSR}$ begins to increase. 
In Fig.~\ref{fig:vlsr_L_MS}, we present LSR velocity centroids of our \ion{C}{4} (half-top circles) and \ion{Si}{4} MS absorbers (half-bottom circles) by MS longitude, along with \ion{C}{4} and \ion{Si}{4} absorbers from \cite{fox20} (half-top and half-bottom diamonds, respectively) and \ion{O}{6} absorbers from \cite{sembach03} (x marks). For \ion{O}{6} absorbers, we only include the absorbers identified as MS-associated in \cite{sembach03}. The exceptions are absorbers detected in the sight line of H1821+643 ($L_{\rm MS} = -174.1^\circ$, $B_{\rm MS} = -7.4^\circ$), which were newly identified in their work.
We also show \ion{H}{1} contours and a linear extrapolation of the \ion{H}{1} velocity gradient from \cite{nidever10} (dashed line). Our \ion{C}{4} MS absorbers closely follow the \ion{H}{1} velocity gradient, including the velocity inflection, reinforcing their association with the Stream.

We use ionization structure as an additional test for association with the Stream. In several PQS sight lines, we also detect \ion{Si}{4} (Fig.~\ref{fig:vpf}), thus we compare the \ion{Si}{4}/\ion{C}{4} column density ratios with those measured in previously studied Stream sight lines. 
In Fig.~\ref{fig:SiIV_CIV}, we present the spatial distribution of \ion{Si}{4}/\ion{C}{4} (left) for both our MS absorbers (filled) and those from \citet{fox14} (open). Triangles denote upper limits in cases where \ion{Si}{4} is not detected. The right panel shows the ratio as a function of MS latitude ($B_{\rm MS}$). 
We find that the \ion{Si}{4}/\ion{C}{4} ratio decreases with decreasing $B_{\rm MS}$, and our new PQS sight lines fall along the same trend as previously studied MS absorbers. While \ion{Si}{4}/\ion{C}{4} is not a straightforward tracer of ionization state, it is noteworthy that the ratio exhibits a dependence on MS latitude. Previous studies have explored ion ratio trends with MS longitude, particularly in low-to-high ion ratios such as \ion{C}{2}/\ion{C}{4} and \ion{Si}{2}/\ion{Si}{4}. These ratios decrease toward larger negative $L_{\rm MS}$, suggesting that the gas becomes more ionized toward the Stream’s trailing end \citep{fox14, kim24}. 
This newly found trend with MS latitude provides additional support that our absorbers are part of the Stream. A detailed exploration of its physical meaning is left to future work with full ionization modeling, but this result may offer a new diagnostic for understanding the origin of the ionized Stream.

Despite kinematic alignment and a consistent ionization trend with the Stream, potential contamination sources must be considered, particularly other Galactic halo HVCs such as the Outer Arm and Complex~C (Fig.~\ref{fig:targets_dist}). 
Since several of our sight lines intersect regions associated with these known HVCs, it is necessary to assess whether the MS absorbers are genuinely part of the Stream or potentially associated with other HVCs. 
In the sight lines above the Galactic plane (J1938+5408, J1858+4850, J1939+7007, and H1821+643), we detect additional absorption components at $v_{\rm LSR} \sim -130\ \rm km\ s^{-1}$, alongside the MS absorbers (green; Fig.~\ref{fig:vpf}). 
These components are also seen in lower ionization states, including \ion{Al}{2} and \ion{Si}{2}, whereas the MS absorbers are only detected in \ion{C}{4} (Table~\ref{tab:all_VPF}). 
The LSR velocities and ionization states of these components are consistent with the Outer Arm \citep{tripp12}.

Another potential contaminant is the high-velocity ridge of Complex~C, a widespread structure with characteristic velocities around $v_{\rm LSR} \sim -180\ \rm km\ s^{-1}$ \citep{wakker01, tripp03}. 
Notably, the \ion{C}{4} absorption at $v_{\rm LSR} = -215\ \rm km\ s^{-1}$ in the sight line toward H1821+643, identified as an MS absorber here, has previously been reported as a peculiar ionized HVC. 
\cite{savage95} noted that this feature is only detected in \ion{C}{4}, not in any other lower ion species, \ion{H}{1}, \ion{Mg}{2}, \ion{Si}{2}, or \ion{S}{2}. 
Similarly, \cite{tripp03} showed the presence of \ion{C}{4} only and suggested that it may trace an extended, highly ionized part of the high-velocity ridge. 
\cite{lehner10} showed widespread detections of highly ionized HVCs at $v_{\rm LSR} \approx -110$ and $-180\ \rm km\ s^{-1}$ in sight lines near H1821+643, further supporting the existence of a prevalent ionized structure. However, unlike the absorber toward H1821+643, these ionized HVCs generally detected in lower ion species (e.g., \ion{Si}{2}, \ion{Si}{3}, \ion{C}{2}) and tend to have somewhat less negative LSR velocity of $-180$ to $-150\ \rm km\ s^{-1}$. 

In contrast, the MS absorbers above the Galactic plane are detected solely in \ion{C}{4}, indicating a higher ionization state. 
Their LSR velocities span $v_{\rm LSR} = -215$ to $-300\ \rm km\ s^{-1}$, far beyond the range associated with the high-velocity ridge. Moreover, they align closely with \ion{H}{1} velocity gradient of the Stream (Fig.~\ref{fig:vlsr_L_MS}). 
These characteristics support the interpretation that the MS absorbers trace a distinct gaseous structure, likely a highly ionized extension of the Stream, rather than the high-velocity ridge or other Galactic HVCs.

The last potential contamination source is the CGM of M31. 
\cite{lehner20} conducted Project AMIGA, an absorption line survey using QSO sight lines to probe the CGM of M31, and demonstrated that it overlaps with the Stream in both sky position and velocity space. 
To determine absorber associations, they examined the $L_{\rm MS}-v_{\rm LSR}$ plane and adopted the velocity range defined by \cite{nidever10} as a reference for identifying Stream-associated absorption. Their findings indicate that absorbers falling within this velocity range are highly likely to be associated with the Stream, and therefore, support the identification of our newly detected \ion{C}{4} absorbers as Stream-associated.

\subsection{Ionization State} \label{subsec:ioniz}

\begin{figure*}
    \centering
    \begin{subfigure}{0.49\linewidth}
        \includegraphics[width=\linewidth]{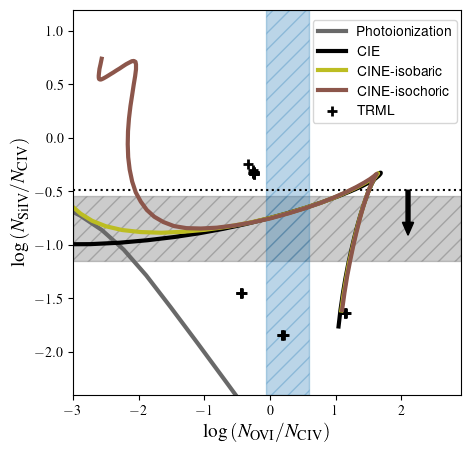}
        \caption{Ion ratios predicted ionization models.}
        \label{fig:ion_ratios}
    \end{subfigure}
    \hfill
    \begin{subfigure}{0.49\linewidth}
        \includegraphics[width=\linewidth]{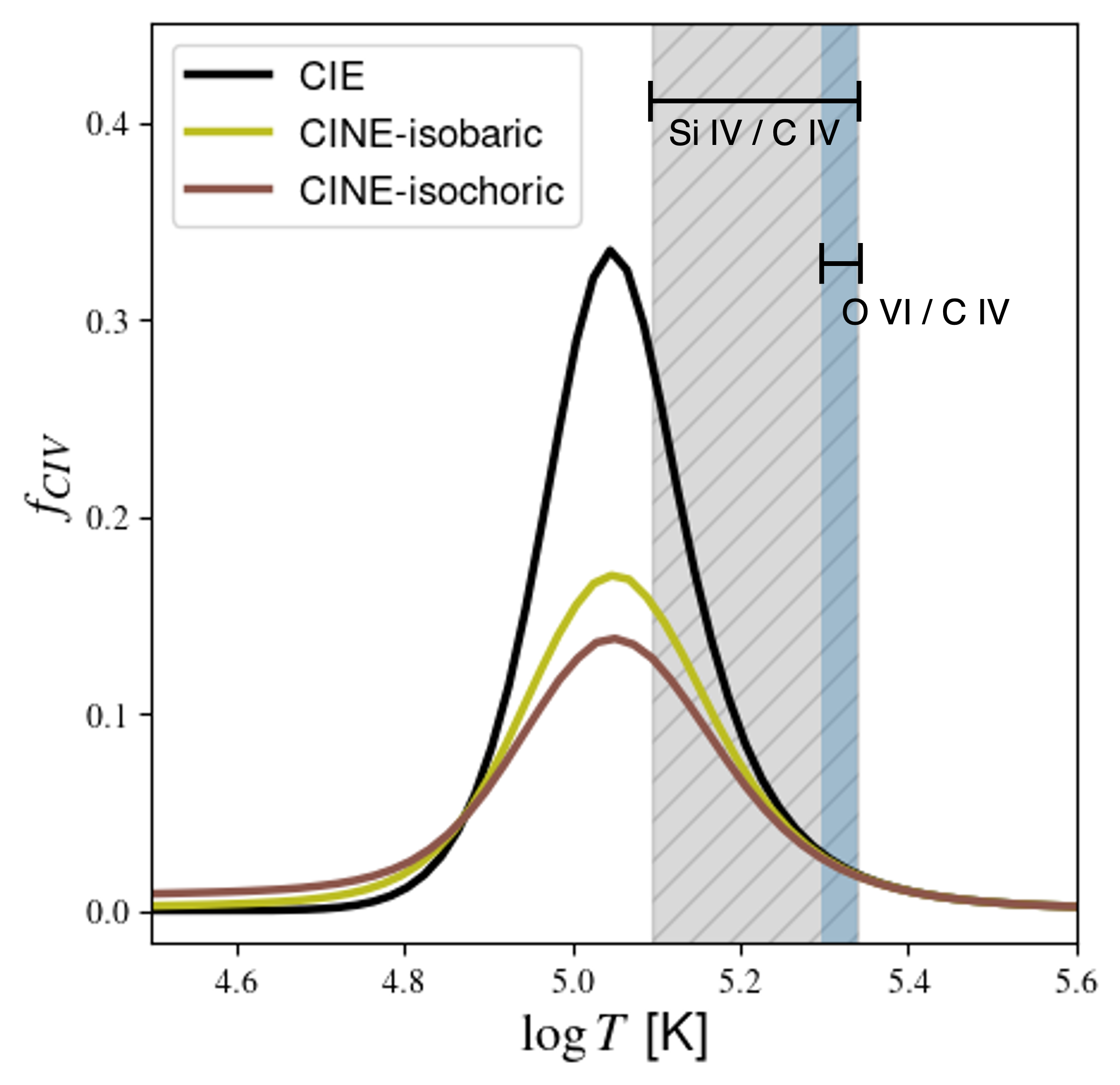}
        \caption{\ion{C}{4} fraction predicted by collisional ionization models.}
        \label{fig:civ_fraction}
    \end{subfigure}
    \caption{(a) Observed ion column density ratios, $N_{\rm SiIV}/N_{\rm CIV}$ (black-hatched) and $N_{\rm OVI}/N_{\rm CIV}$ (blue-hatched), and comparison with ionization models. We present ion ratio predictions from photoionization (gray), collisional ionization equilibrium (CIE; black), radiative cooling (olive and brown), and turbulent radiative mixing layers models (black pluses). 
    (b) The expected \ion{C}{4} fraction from collisional ionization models \citep{gnat07}, collisional ionization equilibrium (black), time-dependent radiative cooling at constant pressure (isobaric; olive) and at constant volume (isochoric; brown). We present sub-solar metallicity ($0.1~Z_{\odot}$) for radiative cooling models. The shaded temperature ranges are derived from observed ion column density ratios, $N_{\rm SiIV}/N_{\rm CIV}$ (black) and $N_{\rm OVI}/N_{\rm CIV}$ (blue), respectively.}
    \label{fig:civ_fraction_combined}
\end{figure*}

The MS absorbers are primarily detected in \ion{C}{4}, with no detection in low ions such as \ion{Si}{2}. It indicates that the gas is highly ionized, consistent with prior studies showing increased ionization toward the trailing end of the Stream based on declining \ion{C}{2}/\ion{C}{4} ratios \citep{kim24}. 

To investigate the ionization mechanism, we consider diagnostic ion ratios such as \ion{Si}{4}/\ion{C}{4} and \ion{O}{6}/\ion{C}{4}, which are commonly used to distinguish between photoionization and collisional ionization scenarios. 
While our dataset contains only \ion{C}{4} and limited \ion{Si}{4} detections, we incorporate measurements of \ion{O}{6} and \ion{Si}{4} from literature along the Stream, particularly for the tip of the Stream at $ L_{\rm MS} < -105^\circ$, to estimate these ratios where possible \citep{sembach03, fox14}.

We performed ionization modeling under both photoionization and collisional ionization scenarios and present the model predictions in Fig.~\ref{fig:ion_ratios}. 
For photoionization, we considered Cloudy models (C17; \citealt{ferland17}) that assume ionization equilibrium and radiation from the Milky Way and the extragalactic background of \cite{haardt96} \citep{fox05}. These models fail to reproduce the observed \ion{O}{6}/\ion{C}{4} ratios, which range from $-0.07 \lesssim \log{(N_{\rm OVI}/N_{\rm CIV})} \lesssim 0.59$, instead predicting $\log{(N_{\rm OVI}/N_{\rm CIV})} < -1$ over a wide range of densities and temperatures (gray; Fig.~\ref{fig:ion_ratios}). 
Although models with extremely low \ion{H}{1} column densities of $N_{\rm HI} < 14.5\ \rm cm^{-2}$ can yield higher \ion{O}{6}/\ion{C}{4} ratios, they simultaneously predict very low \ion{Si}{4}/\ion{C}{4} values, $\log{(N_{\rm SiIV}/N_{\rm CIV})} < -2.5$, which are inconsistent with the observed range of -1.15 to -0.34. 
This result may suggest that a more extreme radiation source is required to explain the observed ionization state with photoionization, if we assume \ion{Si}{4}, \ion{C}{4}, and \ion{O}{6} are in the same phase of gas. 
Alternatively, if we instead assume that the kinematically coincident \ion{O}{6} traces a distinct gas phase, then a photoionization model for the observed \ion{C}{4} and \ion{Si}{4} column densities requires low-density gas with hydrogen number density of $n_{\rm H} < 10^{-4}\ \rm cm^{-3}$ and a temperature of $T \sim 10^{4.3}$~K. We note that, with a typical \ion{C}{4} column density of $N_{\rm CIV} \sim 10^{13.8}\ \rm cm^{-2}$ along the sight lines, this low density implies a very long length scale of the material of $> 100$~kpc along the sight lines, which is unrealistic, especially given that the Stream in this region is predicted to lie at a distance of order $\sim 100$~kpc \citep[e.g.][]{besla12}.

We also apply the analytic model of turbulent radiative mixing layers (TRMLs) developed by \citet{chen23}, adopting a fiducial hot gas pressure of $p_{\rm hot}/k_{\rm B} = 10^4\ \rm K\ cm^{-3}$. We note that the predicted column density ratios from TRMLs remain nearly constant over a broad range of thermal pressures ($p_{\rm hot}/k_{\rm B} = 10^{3-5}\ \rm K\ cm^{-3}$). We explore a wide parameter space in density contrast between the hot and cold phases ($\chi = 5$–1000) and velocity shear ($v_{\rm shear} = 100$–400 km s$^{-1}$). The predicted column density ratios from these models are presented in Fig.~\ref{fig:ion_ratios} (black pluses). 
Some of the TRML models reproduce the observed $N_{\rm OVI}/N_{\rm CIV}$ column density ratios. However, the predicted $N_{\rm SiIV}/N_{\rm CIV}$ ratios are typically lower than the observed values. We note, however, that at the tip of the Stream, where only \ion{C}{4} is detected and \ion{Si}{4} is not, the observed $N_{\rm SiIV}/N_{\rm CIV}$ represents an upper limit. n such cases, the TRML models may plausibly explain the observed ionization conditions.

Finally, we compare the observed ion ratios to predictions from collisional ionization models, including collisional ionization equilibrium (CIE) and non-equilibrium radiative cooling models \citep{gnat07}. These models can simultaneously reproduce the observed $N_{\rm SiIV}/N_{\rm CIV}$ and $N_{\rm OVI}/N_{\rm CIV}$ ratios (Fig.~\ref{fig:ion_ratios}) with gas at a temperature of $\log{T} \approx 5.3$ (Fig.~\ref{fig:civ_fraction}). If \ion{Si}{4}, \ion{C}{4}, and \ion{O}{6} all arise from a single, collisionally ionized phase, these models offer a self-consistent explanation for the observed ionization state.

\subsection{Mass Estimation} \label{subsec:mass} 
The total gas mass in the Stream serves as a critical constraint on the Stream formation and the LMC-SMC and MW dynamics models. 
It also provides important context for the baryon-to-dark matter (DM) ratio of the Clouds, especially the LMC. The rotation curve of the LMC out to $\sim 8.7$~kpc suggests a total mass of $\sim 1.7 \times 10^{10}\ M_\odot$ \citep{vandermarel14}, which yields a high baryon fraction of $ M_{\rm b}/M_{\rm DM}\sim 11 \%$. If a substantial portion of the Stream’s mass was originally stripped from the LMC, this would imply an even higher initial baryon fraction, potentially approaching the cosmic baryon fraction of $\sim 15 \%$ \citep{besla15}.

The \ion{H}{1} mass of the Stream is estimated to be $\sim 2.7 \times 10^8\ M_{\sun}$, assuming a distance of 55~kpc \citep{bruns05}. This corresponds to roughly one-third of the total \ion{H}{1} gas mass in the Clouds, suggesting that a substantial amount of neutral gas has been stripped to form the Stream. 
In addition, the Stream likely contains an even greater mass in ionized form, estimated at $\sim 10^9\ M_{\sun}$ \citep{fox14}. The ionized gas mass may also be important for examining the properties of the MW halo and Magellanic corona \citep{lucchini24a}, as some models suggest that a significant fraction of the ionized component may originate in the Magellanic corona and be incorporated into the Stream through interactions with the MW halo \citep{lucchini20}, though recent simulations by \citet{zhu24} suggest that the CGM of dwarf satellites would be rapidly stripped in MW-like host environments, even for relatively massive dwarfs such as the LMC.

Our discovery of an extended ionized gas adds to the current gas mass estimate. In this section, we estimate the ionized gas mass in this region and discuss its implications for our understanding of the Stream. 

An ionization correction using column densities of ions in different ionization states is required to estimate the total ionized gas mass. We derive the total ionized gas mass $M_{\rm HII}$, following the approach of \cite{fox14}:

\begin{align}
    M_{\rm HII} &= m_{\rm H} A({\rm H\ II})\left<N_{\rm H II}\right> \nonumber \\
    &= m_{\rm H} A({\rm H\ II}) \frac{\left<N_{\rm CIV}\right>}{f_{\rm CIV}[\rm C/H]},
\end{align}
where $m_{\rm H}$ is the hydrogen mass, $A(\rm H\ II)$ is the covering area, and $f_{\rm CIV}$ is the \ion{C}{4} fraction relative to the total carbon gas. 
The angular extent of our newly detected ionized gas, $\approx 1800\ \rm deg^2$, corresponds to a covering area of $A(\rm H\ II) \approx 1.6 \times 10^{46}\ \rm cm^{2}\ (d/55~kpc)^2$. 
Applying the mean \ion{C}{4} column density of the MS absorbers, $<N_{\rm CIV}> = 5.19 \times 10^{13}\ \rm cm^{-2}$, the total ionized gas mass can be expressed as follows.

\begin{equation}
    M_{\rm HII} \approx 8.66 \times 10^7\ M_{\odot}\ \left(\frac{d}{55\ \rm kpc}\right)^2 \left(\frac{0.3}{f_{\rm CIV}}\right) \left(\frac{0.1\ Z_{\odot}}{Z}\right), \label{eq:mass}
\end{equation}

We note that, since only \ion{C}{4} is detected in most sight lines, we cannot observationally constrain $f_{\rm CIV}$. 
As discussed in Section~\ref{subsec:ioniz}, ionization models predict that $f_{\rm CIV}$ varies significantly with ionization conditions and temperature. Our ion ratios indicate a value of $f_{\rm CIV}$ ranging from 0.03 to 0.3 (Fig.~\ref{fig:civ_fraction}). The upper limit of 0.3 represents the maximum fraction predicted by both photoionization and collisional ionization models, and therefore provides a conservative lower limit on the total mass. 
Assuming a metallicity of $Z = 0.1\ Z_{\odot}$, we estimate corresponding ionized hydrogen column densities of $\log{<N_{\rm HII}>} = 19.83$ and $18.83\ \rm cm^{-2}$ for $f_{\rm CIV} = 0.03$ and 0.3, respectively. These values are broadly consistent with previous estimates of $\log{N_{\rm HII}} = 18.93 \pm 0.5\ \rm cm^{-2}$ in regions where \ion{H}{1} is not detected \citep{fox14}, and are comparable to predictions from hydrodynamical simulations \citep{lucchini24a}. 
Under these assumptions, the total ionized gas mass in the newly discovered extension could range from 6\% to 60\% of the previously estimated ionized gas mass of the Stream.

\section{Discussion} \label{sec:discussion}
We find that the PQS \ion{C}{4} absorbers discussed in previous sections are most likely associated with the Stream. This discovery reveals that the ionized component of the Stream extends approximately $30^\circ$ beyond the known \ion{H}{1} structure \citep{nidever10}. 
Notably, the absorbers extend to the northern side, crossing the Galactic plane, providing directionality of the Clouds' orbit.

The newly discovered \ion{C}{4} absorbers preserve the Stream's kinematics (Fig.~\ref{fig:vlsr_L_MS}), indicating that the ionized gas has not yet fully mixed with the MW's hot halo. 
Assuming a halo temperature of $T \sim 10^6$~K and a radius of $\sim 200$~kpc, the sound crossing time of the halo is about 1~Gyr. This suggests that the gas may have been stripped from the Clouds relatively recently, within the past $\lesssim 1$~Gyr, before dynamical mixing with the ambient halo could erase its coherent velocity structure.

The detection of \ion{C}{4} implies gas at $T \sim 10^5$ K. Notably, \ion{Si}{4} is not detected at the very trailing end, with $\log{(N_{\rm SiIV}/N_{\rm CIV})} \lesssim -0.8$. 
For the sight lines where \ion{Si}{4} is detected, \ion{Si}{4}/\ion{C}{4} ratio ranges in $-1.0 \leq \log{(N_{\rm SiIV}/N_{\rm CIV})} \leq -0.54$. 
This low ratio is inconsistent with predictions from AGN-type photoionization spectra, such as those invoked in the Seyfert flare scenario, which typically produce $\log{(N_{\rm SiIV}/N_{\rm CIV})} \gtrsim -0.2$ \citep{bland-hawthorn19}. 

The observed ion ratios are consistent with gas at $\log T \approx 5.3$, supporting a scenario where ionization arises from radiative cooling or turbulent mixing during interaction between the Stream and the clumpy, low-latitude Galactic halo. 
At this temperature, the radiative cooling time is estimated to be $\sim 200$~Myr, assuming metallicity of $0.1~Z_{\odot}$ \citep{mcquinn18}. It suggests that the ionized gas could be produced and remain for hundreds of Myr.

The distance to the Stream remains poorly constrained and almost certainly varies along the length of the Stream.
A recent study by \cite{mishra25} establishes a firm lower limit of 20~kpc for the distance to the Stream. 
They further suggest tentative lower limits of 42~kpc and 55~kpc for regions at $L_{\rm MS} \approx -80^\circ$ and $\approx -100^\circ$, respectively. 
Dynamical models predict that the trailing end of the Stream extends to the farthest distances, reaching $d \gtrsim 120~\rm kpc$ \citep{diaz12, besla12}. 
This strongly suggests that a considerable mass may be deposited at the most distant regions of the Stream.

\section{Summary}
We present the discovery of an ionized extension of the Stream that extends to the northern side of the Galactic Plane.  This extension was found with PQS HST/COS G160M observations that target QSO sight lines at low Galactic latitudes ($b < 30^\circ$), targetting largely unexplored regions of the MW CGM. 
This Stream extension reaches approximately 60~degree beyond the previously studied ionized Stream. 
We detect \ion{C}{4} absorption in the PQS sight lines that fall into an extension in position and LSR velocity of the Stream. 
This ionized extension is mainly traced by \ion{C}{4}, with \ion{Si}{4} detections in some sight lines. We do not detect any low ions, implying the gas is highly ionized. 
We suggest these \ion{C}{4} absorbers are associated with the Stream, given that (1)  velocity centroids of \ion{C}{4} absorbers align with the velocity gradient of the Stream derived from \ion{H}{1} gas, (2) column density ratios \ion{Si}{4}/\ion{C}{4} are consistent with those found in sight lines near the traditional end of the Stream, following a similar trend with Magellanic latitude, and (3) the absorbers present distinct velocity and ionization states from known Galactic HVCs. 
We explore ionization models, including photoionization, CIE, radiative cooling, and TRMLs, based on the observed ion ratios of \ion{Si}{4}, \ion{C}{4}, and \ion{O}{6}. We find that this ionized extension is in a very low-density photoionized gas ($n_{\rm H} < 10^{-4}\ \rm cm^{-3}$) if \ion{O}{6} arises from a distinct gas phase or warm gas phase at $T \approx 10^{5.3}$~K under the assumption that \ion{Si}{4}, \ion{C}{4}, and \ion{O}{6} originate from a single gas phase. We estimate that this ionized extension contributes approximately 6-60\% of the previously estimated ionized mass of the Stream, depending on the \ion{C}{4} fraction and distance. 
Given its position above the Galactic plane, this extension likely traces some of the oldest stripped material in the Stream, and its coherent kinematics suggest it was likely stripped and formed within the past few hundred Myr and has not yet mixed with the hot halo.

\begin{acknowledgments}
We are thankful to the anonymous referee for the constructive comments and suggestions on this paper. 
BC and JKW would like to thank Andy Fox, Sapna Mishra, David Nidever, and Erica Chwalik for valuable conversations around the interpretation of our results. 
This research is based on observations made with the NASA/ESA Hubble Space Telescope obtained from the Space Telescope Science Institute, which is operated by the Association of Universities for Research in Astronomy, Inc., under NASA contract NAS 5–26555. Support for this work was provided by NSF-CAREER 2044303 and HST-GO-16679. 
This research has also made use of the HSLA database, developed and maintained at STScI, Baltimore, USA. 
This work used OpenAI’s ChatGPT (GPT-4) to assist in refining Python scripts for data visualization and to provide limited editorial suggestions during manuscript preparation. 
\end{acknowledgments}

%

\vspace{5mm}
\facilities{HST(COS), HST(STIS)}


\software{Astropy \citep{astropy:2013, astropy:2018, astropy:2022},  
          Cloudy (C17; \citealt{ferland17}), 
          linetools \citep{prochaska17a}, 
          pyigm \citep{pyigm}, 
          veeper \citep{burchett24}}

\dataset{ddd}



\appendix

\section{Voigt Profile Fitting} \label{appendix:vpf}

\startlongtable
\begin{deluxetable*}{llccccc}
    \tablewidth{\textwidth}
    \tablecaption{Best-fit Voigt profile fitting parameters to absorption components. \label{appendix:vpf_tab}}
    \tablehead{
        \colhead{QSO} & 
        \colhead{$(L_{\rm MS}, B_{\rm MS})$} & 
        \colhead{Ion Species} & 
        \colhead{$\log{N_{\rm ion}}$} & 
        \colhead{$b$} & 
        \colhead{$v_{\rm LSR}$} & 
        \colhead{Identification} \\ 
        \colhead{} & 
        \colhead{(deg)} & 
        \colhead{} & 
        \colhead{$(\rm cm^{-2})$} & 
        \colhead{($\rm km\ s^{-1}$)} & 
        \colhead{($\rm km\ s^{-1}$)} & 
        \colhead{}}
    \startdata
    J2206+2757 & (-123.0, -10.4) & \ion{Al}{2} & 12.13 $\pm$ 0.52 & 5.0 $\pm$ 20.3 & 49.2 $\pm$ 7.4 &  \\
               &    & \ion{Si}{2} & 13.46 $\pm$ 0.12 & 9.0 $\pm$ 6.0 & 54.1 $\pm$ 3.0 & \\
               &    & \ion{Si}{4} & 13.39 $\pm$ 0.30 & 20.0 $\pm$ 7.8 & 20.6 $\pm$ 6.0 & \\
               &    &             & 13.02 $\pm$ 0.11 & 44.9 $\pm$ 15.4 & -449.7 $\pm$ 10.2 & MS \\
               &    & \ion{C}{4} & 13.95 $\pm$ 1.5 & 26.8 $\pm$ 27.3 & 28.7 $\pm$ 19.6 & \\
               &    &     & 13.78 $\pm$ 0.04 & 45.0 $\pm$ 5.7 & -345.8 $\pm$ 3.9 & MS \\
    \hline
    J2215+2902 & (-123.1, -8.2) & \ion{Al}{2} & 13.79 $\pm$ 0.49 & 27.9 $\pm$ 11.7 & -30.7 $\pm$ 7.8 &  \\ 
               &        &     & 12.39 $\pm$ 3.09 & 4.1 $\pm$ 79.2 & -84.3 $\pm$ 25.4 &  \\
               &                & \ion{Si}{2} & 14.76 $\pm$ 0.15 & 23.3 $\pm$ 4.5  & -13.3 $\pm$ 3.9 & \\
               &         &      & 14.35 $\pm$ 0.12 & 15.9 $\pm$ 4.3  & -63.0 $\pm$ 4.2 & \\
               &                & \ion{Si}{4} & 13.90 $\pm$ 0.04 & 32.7 $\pm$ 3.7  & -16.9 $\pm$ 2.9 & \\
               &        &     & 12.91 $\pm$ 0.25 & 16.9 $\pm$ 13.5 & -80.8 $\pm$ 9.1 & \\
               &        &     & 13.02 $\pm$ 0.12 & 41.1 $\pm$ 16.2 & -429.8 $\pm$ 10.7 & MS \\
               &        & \ion{C}{4}  & 13.67 $\pm$ 0.71 & 10.8 $\pm$ 16.78 & 20.4 $\pm$ 10.3 & \\
               &       &      & 14.44 $\pm$ 0.14 & 30.0 $\pm$ 6.4 & -13.5 $\pm$ 9.1 & \\
               &       &      & 13.50 $\pm$ 0.09 & 30.0 $\pm$ 8.9 &-388.8 $\pm$ 5.8 & MS \\
    \hline
    J2251+3419  & (-124.9, 1.0) & \ion{Al}{2} & 13.43 $\pm$ 0.34 & 34.2 $\pm$ 14.7 & -14.7 $\pm$ 23.0 & \\
                &       &      & 13.54 $\pm$ 0.84 & 17.5 $\pm$ 11.3 & -59.5 $\pm$ 15.2 &  \\
                &               & \ion{Si}{2} & 14.78 $\pm$ 0.21 & 24.9 $\pm$ 6.3 & -5.8 $\pm$ 7.4 & \\
                &        &     & 14.31 $\pm$ 0.22 & 18.1 $\pm$ 7.2 & -53.7 $\pm$ 9.0 &  \\ 
                &              & \ion{Si}{4} & 13.36 $\pm$ 0.16 & 37.2 $\pm$ 11.2 & -21.6 $\pm$ 12.3 & \\
                &        &     & 12.99 $\pm$ 0.38 & 24.6 $\pm$ 12.2 & -73.4 $\pm$ 12.2 & \\
                &              & \ion{C}{4} & 13.81 $\pm$ 0.18 & 35.0 $\pm$ 13.1 & -2.0 $\pm$ 12.7 & \\
                &       &     & 13.60 $\pm$ 0.29 & 31.7 $\pm$ 15.8 & -65.1 $\pm$ 17.0 & \\
                &       &     & 13.73 $\pm$ 0.05 & 41.1 $\pm$ 6.9 & -412.7 $\pm$ 4.8 & MS \\
    \hline
    J2203+3145  & (-126.7, -9.3) & \ion{Al}{2} & 13.93 $\pm$ 0.18 & 38.2 $\pm$ 4.4 & -32.1 $\pm$ 1.7 & \\
                &                & \ion{Si}{2} & 16.47 $\pm$ 1.46 & 23.8 $\pm$ 7.5 & -19.6 $\pm$ 0.9 & \\
                &                & \ion{Si}{4} & 13.58 $\pm$ 0.05 & 22.0 $\pm$ 2.7 & 0.3 $\pm$ 2.3 & \\
                &       &      & 13.29 $\pm$ 0.09 & 30.9 $\pm$ 8.3 & -57.3 $\pm$ 6.3 & \\
                &              & \ion{C}{4} & 14.29 $\pm$ 0.02 & 30.9 $\pm$ 2.1 & 5.8 $\pm$ 1.5 & \\
                &       &      & 13.36 $\pm$ 0.11 & 20.1 $\pm$ 7.8 & -64.9 $\pm$ 5.3 & \\
                &       &     & 13.89 $\pm$ 0.03 & 48.0 $\pm$ 4.2 & -383.2 $\pm$ 2.9 & MS \\
    \hline
    J2141+3151  & (-129.1, -13.3) & \ion{Al}{2} & 13.55 $\pm$ 0.05 & 63.5 $\pm$ 7.3 & -19.6 $\pm$ 5.1 & \\
                &                 & \ion{Si}{2} & 14.84 $\pm$ 0.04 & 63.4 $\pm$ 4.1 & -22.7 $\pm$ 2.8 & \\
                &                 & \ion{Si}{4} & 13.50 $\pm$ 0.04 & 35.0 $\pm$ 4.6 & -10.5 $\pm$ 3.2 & \\
                &       &      & 13.19 $\pm$ 0.09 & 49.6 $\pm$ 13.6 & -378.6 $\pm$ 9.4 & MS \\
                &              & \ion{C}{4}  & 14.07 $\pm$ 0.05 & 40.2 $\pm$ 5.2 & 9.7 $\pm$ 3.7 & \\
                &       &      & 14.16 $\pm$ 0.04 & 53.6 $\pm$ 5.8 & -343.7 $\pm$ 4.1 & MS \\
    \hline
    J2109+3532  & (-136.1, -16.7) & \ion{Al}{2} & 13.01 $\pm$ 0.22 & 31.9 $\pm$ 15.4 & 73.2 $\pm$ 15.6 & \\
                &       &      & 14.44 $\pm$ 0.82 & 40.0 $\pm$ 15.2 & -31.3 $\pm$ 7.1 & \\
                &                 & \ion{Si}{2} & 14.54 $\pm$ 0.19 & 22.8 $\pm$ 8.0 & 70.1 $\pm$ 6.7 & \\
                &       &      & 15.43 $\pm$ 0.54 & 38.0 $\pm$ 11.3 & -29.1 $\pm$ 6.2 & \\
                &              & \ion{Si}{4} & 12.80 $\pm$ 0.43 & 25.0 $\pm$ 28.9 & 67.6 $\pm$ 21.6 & \\
                &       &      & 13.36 $\pm$ 0.13 & 32.1 $\pm$ 12.8 & 1.4 $\pm$ 8.8 & \\
                &       &      & 13.04 $\pm$ 0.11 & 35.0 $\pm$ 16.7 & -346.1 $\pm$ 11.4 & MS \\
                &              & \ion{C}{4} & 13.75 $\pm$ 0.11 & 28.8 $\pm$ 10.9 & 95.0 $\pm$ 7.0 & \\
                &       &      & 13.89 $\pm$ 0.08 & 32.7 $\pm$ 9.4 & 8.9 $\pm$ 6.1 & \\
                &       &      & 13.90 $\pm$ 0.07 & 30.0 $\pm$ 6.3 & -321.8 $\pm$ 4.5 & MS \\
    \hline
    J1938+5408  & (-161.3, -13.6) & \ion{Al}{2} & 14.10 $\pm$ 0.70 & 39.0 $\pm$ 15.5 & -41.7 $\pm$ 28.6 & \\
                &       &      & 13.60 $\pm$ 0.71 & 27.2 $\pm$ 16.4 & -128.2 $\pm$ 29.1 & OA? \\
                &                 & \ion{Si}{2} & 15.98 $\pm$ 0.66 & 29.78 $\pm$ 6.2 & -32.8 $\pm$ 2.5 & \\
                &       &      & 14.75 $\pm$ 0.15 & 22.8 $\pm$ 3.9 & -128.9 $\pm$ 2.6 & OA? \\
                &              & \ion{Si}{4} & 13.07 $\pm$ 0.16 & 32.8 $\pm$ 13.5 & -15.1 $\pm$ 10.4 & \\
                &       &      & 13.64 $\pm$ 0.05 & 46.0 $\pm$ 7.1 & -101.8 $\pm$ 5.0 & OA? \\
                &              & \ion{C}{4}  & 13.66 $\pm$ 0.04 & 49.5 $\pm$ 6.4 & -16.1 $\pm$ 4.6 & \\
                &       &      & 13.68 $\pm$ 0.04 & 36.0 $\pm$ 3.6 & -111.3 $\pm$ 2.8 & OA? \\
                &       &      & 13.37 $\pm$ 0.03 & 25.3 $\pm$ 3.0 & -273.9 $\pm$ 1.9 & MS \\
    \hline
    J1858+4850  & (-164.9, -21.0) & \ion{Al}{2} & 13.96 $\pm$ 0.12 & 29.9 $\pm$ 2.4 & -24.7 $\pm$ 1.2 & \\
                &       &      & 13.44 $\pm$ 0.03 & 28.4 $\pm$ 1.9 & -112.7 $\pm$ 1.3 & OA? \\
                &                 & \ion{Si}{2} & 15.48 $\pm$ 0.61 & 26.4 $\pm$ 6.9 & -23.1 $\pm$ 2.6 & \\
                &       &      & 14.75 $\pm$ 0.11 & 26.7 $\pm$ 4.2 & -112.9 $\pm$ 2.8 & OA? \\
                &              & \ion{Si}{4} & 13.63 $\pm$ 0.05 & 33.0 $\pm$ 4.4 & -23.3 $\pm$ 3.8 & \\
                &       &      & 13.48 $\pm$ 0.08 & 37.0 $\pm$ 7.2 & -98.5 $\pm$ 5.9 & OA? \\
                &              & \ion{C}{4}  & 14.17 $\pm$ 0.05 & 29.7 $\pm$ 4.2 & -12.8 $\pm$ 3.4 & \\
                &       &      & 14.00 $\pm$ 0.07 & 31.2 $\pm$ 5.9 & -83.5 $\pm$ 4.9 & OA? \\
                &       &      & 13.35 $\pm$ 0.09 & 11.3 $\pm$ 6.1 & -265.4 $\pm$ 3.2 & MS \\
    \hline
    J1939+7007  & (-169.0, 0.5) & \ion{Al}{2} & 14.37 $\pm$ 0.51 & 41.4 $\pm$ 11.5 & -72.3 $\pm$ 16.9 & OA? \\
                &               & \ion{Si}{2} & 15.81 $\pm$ 0.94 & 28.2 $\pm$ 8.6 & -49.4 $\pm$ 14.1 & \\
                &       &      & 14.50 $\pm$ 0.47 & 29.2 $\pm$ 13.8 & -119.1 $\pm$ 24.6 & OA \\
                &              & \ion{Si}{4} & 13.27 $\pm$ 0.09 & 41.7 $\pm$ 9.5 & -30.6 $\pm$ 7.7 & \\
                &       &      & 13.40 $\pm$ 0.07 & 39.4 $\pm$ 7.1 & -117.7 $\pm$ 5.7 & OA \\
                &              & \ion{C}{4}  & 13.82 $\pm$ 0.16 & 46.6 $\pm$ 15.6 & -45.4 $\pm$ 14.7 & \\
                &      &       & 13.84 $\pm$ 0.15 & 35.6 $\pm$ 8.7 & -118.3 $\pm$ 8.9 & OA \\
                &       &      & 13.02 $\pm$ 0.14 & 10.0 $\pm$ 10.3 & -247.8 $\pm$ 5.1 & MS \\
    \hline
    H1821+643   & (-174.1, -7.4) & \ion{Al}{2} & 16.14 $\pm$ 0.22 & 16.69 $\pm$ 0.79 & -25.0 $\pm$ 1.0 & \\
                &       &      & 14.89 $\pm$ 1.53 & 4.3 $\pm$ 2.3 & -86.9 $\pm$ 1.6 & OA? \\
                &        &     & 13.11 $\pm$ 0.04 & 18.9 $\pm$ 1.9 & -132.1 $\pm$ 1.3 & OA \\
                &                & \ion{Si}{2} & 13.90 $\pm$ 0.13 & 4.9 $\pm$ 0.8 & 34.9 $\pm$ 0.4 & \\
                &       &      & 16.22 $\pm$ 0.17 & 12.6 $\pm$ 2.1 & -8.6 $\pm$ 4.3 & \\
                &       &      & 14.42 $\pm$ 0.22 & 20.0 $\pm$ 10.7 & -52.6 $\pm$ 7.5 & \\
                &       &      & 14.16 $\pm$ 0.10 & 10.0 $\pm$ 1.3 & -89.2 $\pm$ 1.5 & OA? \\
                &       &      & 14.79 $\pm$ 0.14 & 10.8 $\pm$ 0.6 & -133.3 $\pm$ 0.3 & OA \\
                &              & \ion{C}{2}  & 15.40 $\pm$ 0.53 & 8.2 $\pm$ 1.3 & 30.2 $\pm$ 1.2 & \\
                &       &      & 18.05 $\pm$ 0.25 & 28.5 $\pm$ 1.2 & -71.1 $\pm$ 1.3 & OA? \\
                &              & \ion{S}{2} & 15.73 $\pm$ 0.03 & 17.9 $\pm$ 0.5 & -16.1 $\pm$ 0.3 & \\
                &        &     & 14.17 $\pm$ 0.07 & 9.0 $\pm$ 2.2 & -60.6 $\pm$ 1.4 & \\
                &        &     & 14.14 $\pm$ 0.05 & 5.8 $\pm$ 1.3 & -91.6 $\pm$ 0.8 & OA? \\
                &        &    & 14.15 $\pm$ 0.08 & 17.30 $\pm$ 3.6 & -130.7 $\pm$ 2.1 & OA \\
                &             & \ion{Si}{4} & 12.85 $\pm$ 0.05 & 14.56 $\pm$ 2.1 & 28.8 $\pm$ 1.4 & \\
                &       &      & 13.42 $\pm$ 0.03 & 30.2 $\pm$ 2.5 & -23.5 $\pm$ 1.2 & \\
                &       &      & 13.22 $\pm$ 0.04 & 15.6 $\pm$ 1.4 & -79.8 $\pm$ 0.9 & OA? \\
                &       &      & 13.33 $\pm$ 0.02 & 24.1 $\pm$ 1.5 & -126.9 $\pm$ 1.0 & OA \\
                &              & \ion{C}{4}  & 13.16 $\pm$ 0.07 & 14.6 $\pm$ 3.1 & 28.5 $\pm$ 2.0 & \\
                &       &      & 14.07 $\pm$ 0.02 & 12.7 $\pm$ 0.7 & -8.6 $\pm$ 0.4 & \\
                &       &      & 13.91 $\pm$ 0.04 & 29.6 $\pm$ 2.7 & -70.0 $\pm$ 1.8 & OA? \\
                &       &      & 13.82 $\pm$ 0.04 & 26.8 $\pm$ 2.2 & -128.1 $\pm$ 2.1 & OA \\
                &       &      & 13.31 $\pm$ 0.04 & 6.6 $\pm$ 0.9 & -216.2 $\pm$ 0.6 & MS \\
                &              & \ion{N}{5}  & 12.99 $\pm$ 0.09 & 5.5 $\pm$ 2.2 & 19.1 $\pm$ 1.4 & \\
                &       &      & 14.20 $\pm$ 0.03 & 10.4 $\pm$ 0.6 & -8.0 $\pm$ 0.4 & \\
                &        &     & 13.13 $\pm$ 0.10 & 25.0 $\pm$ 6.9 & -104.8 $\pm$ 4.8 & OA \\
    \hline
    \enddata
    \tablecomments{Voigt profile fit parameters for all detected ion species in the PQS sight lines. The Magellanic Stream coordinates, $(L_{\rm MS}, B_{\rm MS})$, of each PQS sight line are also provided.  For each absorption component, we present the ion column density ($N_{\rm ion}$), Doppler $b$-parameter, and velocity centroid in the LSR. Absorption components with velocity offsets distinguishable from the ISM are labeled by their likely association: MS = Magellanic Stream; OA= Outer Arm; C = Complex~C).}
    \label{tab:all_VPF}
\end{deluxetable*}


\bibliography{reference}{}
\bibliographystyle{aasjournalv7}



\end{document}